\documentclass[letter, longauth, traditabstract]{aa} 

\usepackage{comment}
\usepackage{longtable,lscape}
\usepackage{natbib}
\usepackage{graphicx}
\usepackage{answers}
\usepackage{verbatim}
\usepackage{txfonts}
\def\ms{\hbox{\,m\,s$^{-1}$}}         
\def\m2s2{\hbox{\,m$^{2}$\,s$^{-2}$}} 
\def\kms{\hbox{\,km\,s$^{-1}$}}       
\def\Msun{\hbox{$\mathrm{M}_{\odot}$}}             
\def\Rsun{\hbox{$\mathrm{R}_{\odot}$}}
\def\Mjup{\hbox{$\mathrm{M}_{\rm Jup}$}}
\def\Rjup{\hbox{$\mathrm{R}_{\rm Jup}$}}

\def\mp{M_{\rm p}}
\def\rp{R_{\rm p}}

\begin{document}

   \title{The GAPS Programme with HARPS-N@TNG \\ 
          VI: The Curious Case of TrES-4b
          \thanks{Based on observations made with the Italian Telescopio 
                  Nazionale Galileo (TNG) operated on the island of La Palma 
                  by the Fundacion Galileo Galilei of the INAF 
                  at the Spanish 
                  Observatorio del Roque de los Muchachos of the IAC
                  in the frame of the program 
                  Global Architecture of Planetary Systems (GAPS), and 
                  with the Zeiss 1.23-m telescope at the German-Spanish Astronomical Center at Calar Alto, Spain
                  }}
\authorrunning{A. Sozzetti et al.}
\titlerunning{GAPS VI: The Curious case of TrES-4b}

   \author{A. Sozzetti    \inst{1},
A.S. Bonomo    \inst{1},
K. Biazzo      \inst{2},
L. Mancini     \inst{3}, 
M. Damasso     \inst{1},
S. Desidera    \inst{4},
R. Gratton     \inst{4},
A.F. Lanza     \inst{2},
E. Poretti     \inst{5},
M. Rainer      \inst{5},
L. Malavolta   \inst{4},
L. Affer       \inst{6},
M. Barbieri    \inst{4},
L.R. Bedin       \inst{4},
C. Boccato    \inst{4},
M. Bonavita    \inst{4},
F. Borsa       \inst{5},
S. Ciceri      \inst{3},
R.U. Claudi    \inst{4},
D. Gandolfi    \inst{2,7},
P. Giacobbe    \inst{1},
T. Henning      \inst{3},
C. Knapic       \inst{8},
D.W. Latham     \inst{9},
G. Lodato      \inst{10},  
A. Maggio      \inst{6},
J. Maldonado   \inst{6},
F. Marzari     \inst{11,4},
A.F. Martinez Fiorenzano \inst{12}, 
G. Micela      \inst{6},
E. Molinari    \inst{12,13},
C. Mordasini   \inst{3},
V. Nascimbeni  \inst{4},
I. Pagano      \inst{2},
M. Pedani      \inst{12},
F. Pepe        \inst{14},
G. Piotto      \inst{11,4},
N. Santos      \inst{15,16},
G. Scandariato \inst{2},
E. Shkolnik \inst{17}, \and
J. Southworth  \inst{18}
           }

\institute{INAF -- Osservatorio Astrofisico di Torino, Via Osservatorio 20, I-10025, Pino Torinese, Italy
\and INAF -- Osservatorio Astrofisico di Catania, Via S.Sofia 78, I-95123, Catania, Italy
\and Max-Planck-Institut f\"ur Astronomie, K\"onigstuhl 17, D-69117, Heidelberg, Germany
\and INAF -- Osservatorio Astronomico di Padova, Vicolo Osservatorio 5, I-35122, Padova, Italy
\and INAF -- Osservatorio Astronomico di Brera, Via E. Bianchi 46, I-23807 Merate (LC), Italy
\and INAF -- Osservatorio Astronomico di Palermo, Piazza del Parlamento, Italy 1, I-90134, Palermo, Italy
\and Landessternwarte K\"onigstuhl, Zentrum f\"ur Astronomie der Universitat Heidelberg, K\"onigstuhl 12, D-69117 Heidelberg, Germany
\and INAF -- Osservatorio Astronomico di Trieste, via Tiepolo 11, I-34143 Trieste, Italy
\and Harvard-Smithsonian Center for Astrophysics, 60 Garden Street, Cambridge, Massachusetts, 02138, USA
\and Dipartimento di Fisica, Universit\`a  di Milano, Via Celoria 16, I-20133 Milano, Italy 
\and Dip. di Fisica e Astronomia Galileo Galilei -- Universit\`a di Padova, Vicolo dell'Osservatorio 2, I-35122, Padova, Italy 
\and Fundaci\'on Galileo Galilei - INAF, Rambla Jos\'e Ana Fernandez P\'erez 7, E-38712 Bre\~na Baja, TF - Spain
\and INAF -- IASF Milano, via Bassini 15, I-20133 Milano, Italy
\and D\'epartement d'Astronomie de l'Universit\'e de Geneve, 51 ch. des Maillettes - Observatoire de Sauverny, CH-1290 Versoix, Switzerland
\and Instituto de Astrof{\'\i}sica e Ci\^encias do Espa\c{c}o, Universidade do Porto, Rua das Estrelas, 4150-762 Porto, Portugal
\and Departamento de F{\'\i}sica e Astronomia, Faculdade de Ci\^encias,  Universidade do Porto, Rua do Campo Alegre, 4169-007 Porto, Portugal
\and Lowell Observatory, 1400 W. Mars Hill Road, Flagstaff, AZ, 86001, USA
\and Astrophysics Group, Keele University, Staffordshire, ST5 5BG, UK
             }

   \date{Received ??; Accepted ??}

 \abstract{We revisit the TrES-4 system parameters based on high-precision HARPS-N radial-velocity measurements and new photometric light curves. 
 A combined spectroscopic and photometric analysis allows us to determine a spectroscopic orbit with an amplitude $K=51\pm3$ m s$^{-1}$. The derived 
 mass of TrES-4b is found to be $M_{\rm p} = 0.49\pm0.04~\Mjup$, significantly lower than previously reported. Combined with the large radius 
 ($R_{\rm p} = 1.84_{-0.09}^{+0.08}~\Rjup$) inferred from our analysis, TrES-4b becomes the second-lowest density transiting hot Jupiter known. 
 We discuss several scenarios to explain the puzzling discrepancy in the mass of TrES-4b in the context of the exotic class of highly inflated 
 transiting giant planets. 
}


\maketitle
%

\section{Introduction}

The class of transiting extrasolar planets (to-date, over 1000 are either confirmed or validated) 
allows for many a study to further our understanding of their interiors, atmospheres, and 
ultimately formation and evolution history (see, e.g. Madhusudhan et al. 2014 and Baraffe et al. 2014). The subset of close-in giant planetary 
companions (hot Jupiters) with very large radii, and corresponding very low mean densities, posed for a time 
a conundrum to theoreticians (the so-called radius anomaly problem; see e.g. Bodenheimer et al. 2003). There is now a growing consensus that 
the radius of a hot Jupiter can be inflated due to several factors, including variable stellar irradiation, 
the planet's mass and heavy element content, tidal and kinetic heating, and Ohmic dissipation (for a review see, e.g., Spiegel et al. 2014, and references therein).

The distribution of planetary radii of transiting hot Jupiters in systems with well-determined stellar and 
planetary parameters has been described in the recent past in terms of some of the relevant factors 
(such as equilibrium temperature, stellar metallicity, and orbital semi-major axis) using empirical formulae based 
on the assumption of independent variables (B\'eky et al. 2011; Enoch et al. 2011) or a multivariate regression approach 
(Enoch et al. 2012; Weiss et al. 2013). 
These latest models are quite successful in statistically reproducing the observed radius distribution of this class of exoplanets.
Still, some of the most extreme planets with the largest radii remain challenging for current models of planetary formation and bulk structure. 
For example, the extremely low densities of objects such as WASP-17b (Anderson et al. 2010), HAT-P-32b (Hartman et al. 2011), WASP-79b (Smalley et al. 2012), 
WASP-88b (Delrez et al. 2014) or Kepler-12b (Fortney et al. 2011) cannot be reproduced by simple models of core-less planets 
(e.g., Baraffe et al. 2014), nor can the atmospheric inflation mechanisms mentioned above explain the observed radii.

TrES-4b (Mandushev et al. 2007, M07 thereafter) is another highly bloated transiting hot Jupiter. It belongs to the restricted lot 
of some dozen objects with a measured radius larger than 1.7 $R_\mathrm{Jup}$ (Sozzetti et al. 2009, S09 thereafter; Chan et al. 2011; Sada et al. 2012; Southworth 2012). 
Measurements of the Rossiter-McLaughlin effect (Narita et al. 2010, N10 hereafter) revealed close spin-orbit alignment of the TrES-4 system. 
Atmospheric characterization measurements have been obtained by Knutson et al. (2009), who detected a temperature inversion in the TrES-4b's broadband infrared 
emission spectrum with Spitzer/IRAC during secondary eclipse, and by Ranjan et al. (2014), who presented a featureless transmission spectrum of TrES-4b 
using HST/WFC3 during primary transit. Constraints from the secondary eclipse measurements and expanded radial-velocity (RV) datasets (Knutson et al. 2014, 
K14 thereafter) indicate a probable circular orbit for TrES-4b. Using a time baseline in excess of five years K14 did not detect any significant 
acceleration in the RV data that might point to the presence of a massive outer companion in the system. Finally, TrES-4 has a faint common proper motion 
companion at $\sim1.5^{\prime\prime}$, discovered by Daemgen et al. (2009) and confirmed by Bergfors et al. (2013). 

In this Letter we present RV measurements of TrES-4 gathered with the HARPS-N spectrograph (Cosentino et al. 2012) on the 
Telescopio Nazionale Galileo (TNG) within the context of the programme Global Architecture of Planetary Systems (GAPS, Covino et al. 2013; Desidera et al. 2013), 
along with additional photometric light-curves during transit obtained with the Zeiss 1.23-m telescope at the German-Spanish Calar Alto Observatory (CAHA). 
A combined analysis allows us to derive a much lower mass for TrES-4b than previously reported, making it the second-lowest density transiting hot Jupiter known to-date.

%

\section{Spectroscopic and Photometric Observations}

The TrES-4 system was observed with HARPS-N on 17 individual epochs between March 2013 and July 2014. The Th-Ar simultaneous
calibration was not used to avoid contamination of the stellar spectrum by the lamp lines (which would affect a proper spectral analysis). 
In addition, the magnitude of the instrumental drift during a night ($\lesssim 1$ m s$^{-1}$) is considerably lower than the typical photon-noise RV errors 
($\simeq 9$ m s$^{-1}$), thus of no impact for faint stars such as TrES-4 (see, e.g., Bonomo et al. 2014; Damasso et al. 2015). 
The reduction of the spectra and the RV measurements were obtained using the latest version (Nov. 2013) of the HARPS-N 
instrument data reduction software (DRS) pipeline and the G2 mask. We measured the RVs using the weighted cross-correlation function (CCF) method 
(Baranne et al. 1996; Pepe et al. 2002). The individual measurements are reported in Table~\ref{table_rv}, 
together with the values of bisector span and chromospheric activity index $\log R^{\prime}_{HK}$.

The spectra of TrES-4 were coadded to produce a merged spectrum with a peak signal-to-noise ratio of $\sim110$ pixel$^{-1}$
at 550 nm. We determined the atmospheric stellar parameters using
the code MOOG (Sneden 1973; version 2013) and implemented both the methods based on equivalent widths and on spectral synthesis, 
as described in Biazzo et al. (2012), D'Orazi et al. (2011), and Gandolfi et al. (2013). 
The final adopted parameters are listed in Table~\ref{starplanet_param_table}.

\begin{table}
\centering
\caption{HARPS-N radial velocities, formal errors, bisector spans, and chromospheric activity index of TrES-4.}
\vspace{0.1cm}
\setlength{\tabcolsep}{1.0mm}
\renewcommand{\footnoterule}{}                          
\begin{tabular}{c c c c c}        
\hline \hline
$ \rm BJD_{TDB}$ & RV & $\pm 1~\sigma$ & Bis. span & $\log R^{\prime}_{HK}$ \\
$-$2450000   & ($\kms$) & ($\kms$) & ($\kms$) & \\
\hline
6362.736988  &  $-$16.069 & 0.007 & $-$0.005&$-$5.046 \\
6484.535732  &  $-$16.139 & 0.008 &  0.034&$-$5.290 \\
6485.620551  &  $-$16.112 & 0.011 & $-$0.011&$-$5.303 \\
6506.435512  & $-$16.143  & 0.007 &  0.018&$-$5.154 \\
6507.544966  &  $-$16.078 & 0.018 &  0.050&$-$5.181 \\
6508.534111  &  $-$16.071 & 0.010 &  $-$0.045&$-$5.259   \\
6509.530003  & $-$16.139  & 0.009 &  $-$0.022&$-$5.257 \\
6543.363868  &  $-$16.052 & 0.007 &  $-$0.010&$-$5.068  \\
6583.321741  &  $-$16.036 & 0.010 & $-$0.041&$-$5.189 \\
6586.336543  &  $-$16.039 & 0.009 &  $-$0.024&$-$5.109  \\
6696.775058  &  $-$16.078 & 0.020 &  $-$0.033&$-$5.058  \\
6699.728820  &  $-$16.049 & 0.009 &  $-$0.012&$-$5.117 \\
6701.762988  &  $-$16.145 & 0.009 &  0.007&$-$5.099  \\
6786.659718  &  $-$16.135 & 0.012 &  $-$0.010&$-$5.160  \\
6787.693684  &  $-$16.130 & 0.010 &  0.010&$-$5.111   \\
6857.483512  & $-$16.109  & 0.007 &  0.030&$-$5.089  \\
6858.451440  & $-$16.151 & 0.008 &   0.027& $-$5.169  \\
\hline \hline
\label{table_rv}
\end{tabular}
\end{table}

We carried out $I_c$-band precision photometric observations of two complete
transit events of TrES-4\,b with the CAHA 1.23-m on UT 2013 July 6 and UT 2014 June 30. 
The telescope was defocussed and autoguided during all the
observations and the CCD was windowed to reduce the readout time. 
The datasets were reduced using standard calibration techniques
(overscan correction, trimming, bias subtraction, flat fielding). We then derived differential fluxes
relative to an ensemble of local comparison stars (using the methodology described in Southworth et al. 2014).
The final set of photometric time series of TrES-4 is available in a machine-readable form in the electronic version of Table~\ref{tbl:phottres4}. 
Uncertainties on individual photometric measurements were estimated separately for the two light curves as the 
standard deviation of the residuals of the transit fitting; these uncertainties are larger than the formal error bars in both cases. 
Correlated noise was then estimated following Pont et al. (2006) and Bonomo et al. (2012) and added in quadrature with the individual measurement uncertainties.
Final uncertainties are equal to $8.42\times10^{-4}$ and $7.71\times10^{-4}$ (in units of relative flux) for the former and latter light curve, respectively.

\begin{table}
\centering
\caption{System parameters of TrES-4. Errors and upper limits refer to $1~\sigma$ uncertainties.}            
\vspace{0.1cm}
\begin{minipage}[t]{9.0cm} 
\setlength{\tabcolsep}{1.0mm}
\renewcommand{\footnoterule}{}                          
\begin{tabular}{l l}        
\hline\hline
\emph{Stellar parameters} &  \\
Effective temperature $T_{\rm{eff}}$[K]& 6295 $\pm$ 65 \\
Metallicity $[\rm{Fe/H}]$ [dex] & 0.28  $\pm$ 0.09 \\
Microturbulence velocity $\xi_{t}$ [\kms] & $1.73 \pm 0.02$ \\
Rotational velocity $V \sin{i_{*}}$ [\kms] & 8.5 $\pm$ 0.5 \\
Systemic velocity $\gamma$ [\ms] & $-16097.0 \pm 2.6$ \\
RV jitter [\ms] & $< 6$ \\
Density $\rho_{*}$ [$ \rm g\;cm^{-3}$] & $0.347_{-0.031}^{+0.042}$ \\
Mass [\Msun] & $1.45 \pm 0.05$ \\
Radius [\Rsun] & $1.81 \pm 0.08$  \\
Derived surface gravity log\,$g$ [cgs] & $4.09 \pm 0.03$ \\
Age $t$ [Gyr] & $2.2 \pm 0.4$ \\
\hline
\emph{Transit and orbital parameters} &    \\
Orbital period $P$ [days] & 3.55392771 (47) \\ 
Transit epoch $T_{ \rm 0} [\rm BJD_{TDB}-2450000$] & 4230.90560 (30) \\ 
$e~\cos{\omega}$ & $ 0.0010_{-0.0017}^{+0.0022}$ \\
$e~\sin{\omega}$ & $ 0_{-0.022}^{+0.012} $ \\
Orbital eccentricity $e$  & $< 0.016 $  \\
Argument of periastron $\omega$ [deg] & unconstrained  \\
RV semi-amplitude $K$ [\ms] & $51.1 \pm 3.3 $ \\
Transit duration $T_{\rm 14}$ [h] & $3.658_{-0.030}^{+0.036}$ \\
Radius ratio $R_{\rm p}/R_{*}$ & $0.10452_{-0.00072}^{+0.00066}$  \\
Inclination $i$ [deg] & $83.07_{-0.44}^{+0.51}$ \\
Linear limb-darkening coefficient $u$  & $0.524_{-0.065}^{+0.060}$ \\
$a/R_{*}$ & $6.14_{-0.19}^{+0.24}$ \\
Impact parameter $b$ & $0.744_{-0.022}^{+0.016}$ \\
\hline
\emph{Planetary parameters} &  \\
Mass $M_{\rm p} ~[\Mjup]$  & $0.494 \pm 0.035$ \\
Radius $R_{\rm p} ~[\Rjup]$  & $1.838_{-0.090}^{+0.081}$ \\
Density $\rho_{\rm p}$ [$\rm g\;cm^{-3}$] & $0.099_{-0.013}^{+0.016} $ \\
Surface gravity log\,$g_{\rm p }$ [cgs] & $2.45 \pm 0.05$ \\
Orbital semi-major axis $a$ [au] & $0.0516 \pm 0.0005$  \\
Equilibrium temperature $T_{\rm eq}$ [K] ~$^b$ & $1795_{-39}^{+35}$\\
\hline\hline       
\vspace{-0.5cm}
\footnotetext[1]{\scriptsize from \citet{Mandushevetal2007}} \\
\footnotetext[2]{\scriptsize Black-body equilibrium temperature assuming a null Bond albedo and uniform 
heat redistribution to the night side.} \\
\end{tabular}
\end{minipage}
\label{starplanet_param_table}  
\end{table}



\begin{figure}[t]
\centering
\includegraphics[width=7cm, angle=90]{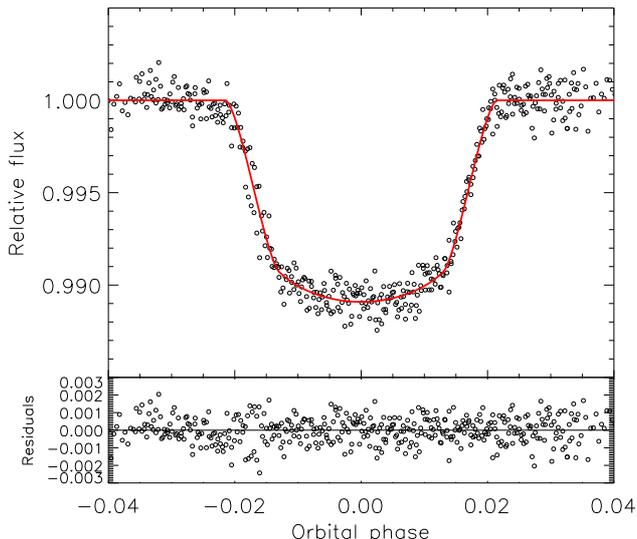}
\vspace{-0.1cm}
\caption{Phase-folded observations of two full transits of TrES-4b in $I_{\rm c}$ band
along with the best-fit model (red solid line). See Section 3 for details.
}
\label{fig_compare_k_values}
\end{figure}

\begin{table}
\centering
\caption{Differential photometry of \mbox{TrES-4}. The full dataset will be made available at the CDS.}            
\setlength{\tabcolsep}{1.0mm}
\begin{tabular}{l c c}   
\multicolumn{3}{c}{$I_c$ band (CAHA 1.23-m)} \\
\hline \hline
BJD$_\mathrm{TDB}$ $-$ 2450000 & Relative flux & Uncertainty \\
\hline
    6480.353148\dotfill     &       1.000482     &       0.000842     \\
    6480.355590\dotfill     &       1.000129     &       0.000842     \\
    6480.357108\dotfill     &       0.999889     &       0.000842     \\
    6480.358670\dotfill     &       0.999212     &       0.000842     \\
    6480.360246\dotfill     &       1.000970     &       0.000842     \\
    6480.361822\dotfill     &       1.000361     &       0.000842     \\
\hline    
\label{tbl:phottres4}
\end{tabular}
\end{table}

\section{Revised TrES-4 System Parameters}

New parameters of the TrES-4 system were derived through a Bayesian combined analysis of our photometry in $I_c$ band and HARPS-N
RV measurements by simultaneously fitting a transit model \citep{Gimenez06, Gimenez09} and a Keplerian orbit. 
For this purpose, we used a Differential Evolution Markov Chain Monte Carlo method (\citealt{TerBraak2006, Eastmanetal2013})
with a Gaussian likelihood function (see, e.g., \citealt{Gregory2005}). 
Our global model has eleven free parameters: the transit epoch $T_{\rm 0}$; the orbital period $P$; 
the systemic radial velocity $\gamma$; the radial-velocity semi-amplitude $K$; $e~{\cos{\omega}}$ and $e~{\sin{\omega}}$, 
where $e$  is the eccentricity and $\omega$ the argument of periastron;
an error term added in quadrature to the formal uncertainties to account for possible jitter in the RV measurements 
regardless of its origin, such as instrumental effects, stellar activity, etc.; 
the transit duration from first to fourth contact $T_{\rm 14}$;
the ratio of the planet to stellar radii $R_{\rm p}/R_{*}$;
the inclination $i$ between the orbital plane and the plane of the sky;
and the coefficient $u$ of a linear limb-darkening law. 
We first tried to use a quadratic limb-darkening law but the two coefficients, especially
the quadratic one, were highly unconstrained. This means that 
the precision of our transit light curves does not allow to fit both coefficients. 

 \begin{figure}[]
\centering
\begin{tabular}{c}
\includegraphics[width=7cm, angle=90]{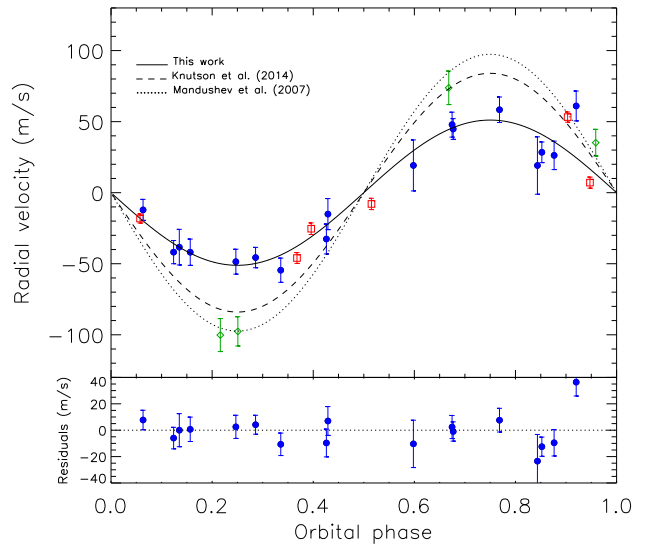} \\
\vspace{0.cm}
\end{tabular}
\caption{\emph{Top panel}: Phase-folded RV measurements of TrES-4 obtained with HARPS-N (blue circles) and, superimposed, the 
best-fit Keplerian orbit model (black solid line). The Keck/HIRES RVs from M07 (green diamonds) and K14 (red squares) and 
the two best-fit orbital solutions obtained in those papers are also shown. 
\emph{Bottom panel}: Residuals from the best-fit model to the HARPS-N radial velocities.}
\label{fig_transit_rv_tres4}
\end{figure}

Gaussian priors were imposed on $T_{\rm 0}$ and $P$, after improving the transit 
ephemeris by combining the transit epochs available in the literature (M07; Chan et al. 2011) 
with the two epochs we derived from our $I_c$ photometry by analyzing 
each individual transit with a circular transit model and a DE-MCMC technique. 
Gaussian priors were also set on the center times of the secondary eclipses observed by 
\citet{Knutsonetal2009} with the Spitzer space telescope because these provide strong
constraints on $e~\cos{\omega}$ (e.g., \citealt{JordanBakos2008}). Non-informative 
priors were used for the other orbital and transit parameters 
while a modified Jeffrey's prior was adopted for the RV jitter term.

The DE-MCMC analysis 
was stopped after reaching convergence and good mixing of the chains
\citep{Ford2006}. The final best-fit transit model and RV curve are overplotted on the phase-folded data in Figures~\ref{fig_compare_k_values} 
and~\ref{fig_transit_rv_tres4}, respectively. We do not determine a significant RV jitter, listed as an upper limit in Table~\ref{starplanet_param_table} 
(only internal errors are reported in Table~\ref{table_rv} and Figure~\ref{fig_transit_rv_tres4}). 
The density of the host star from the transit fitting, and the 
effective temperature and stellar metallicity as derived in Sect.~2 were later compared with the theoretical 
Yonsei-Yale evolutionary tracks \citep{Demarqueetal2004} to determine the 
stellar mass, radius, surface gravity, age, and their associated uncertainties \citep{Sozzettietal2007, Torresetal2012}. 
These are listed in Table~\ref{starplanet_param_table} and 
agree within 1$\sigma$ with the literature values 
(cf., e.g., \citealt{Torresetal2008, Chanetal2011}). 
The related planetary parameters are 
$\mp=0.494 \pm 0.035~\Mjup$, $\rp=1.838_{-0.090}^{+0.081}~\Rjup$, and 
$\rho_{\rm p}=0.099_{-0.013}^{+0.016}~\rm g\;cm^{-3}$. 


\section{Discussion and Conclusions}

Our new, combined spectroscopic and photometric analysis of the TrES-4 system allows us to determine stellar properties in 
good agreement (within the errors) with those measured by M07 and S09. The planetary radius agrees also well 
with the most recent determinations by S09 and Chan et al. (2011), although we formally derive its largest value to-date. 
However, one striking element emerges from our study. 
The best-fit Keplerian orbit for TrES-4 based on HARPS-N RV measurements ($K=51\pm3$ m s$^{-1}$) has an amplitude almost a factor 2 smaller 
than the one ($K=97\pm7$ m s$^{-1}$) reported by M07. As a consequence, the revised mass of the planet is $\sim1.7$ times lower. 
Such a discrepancy clearly deserves a thorough investigation, and we describe here the steps we have taken in this direction. 

In the most recent update of the TrES-4 system parameters, K14 report $K=84\pm10$ m s$^{-1}$ and $\mp=0.843_{-0.089}^{+0.098}$~\Mjup. 
These numbers are compatible within the error-bars with the initial estimates of M07 and S09. The K14 
orbital solution is based on the combination of three datasets, from M07, N10, and obtained by the authors 
themselves. We show in Figure~\ref{fig_transit_rv_tres4} a phase plot of the published Keck velocities and our HARPS-N dataset, superposed on the three orbital 
solutions derived by M07, K14, and in this work (the N10 RV set obtained with Subaru/HDS is of significantly lower internal precision 
and is not shown). The RVs published by M07 are clearly incompatible with the $K$-value derived based on HARPS-N RV data. 
The data obtained by K14 did not sample the critical orbital phases, and they appear consistent with both solutions. 
The larger $K$-value in the Keck data is thus driven by the observations of the discovery paper. 

\begin{figure}[h]
\centering
\includegraphics[width=7cm]{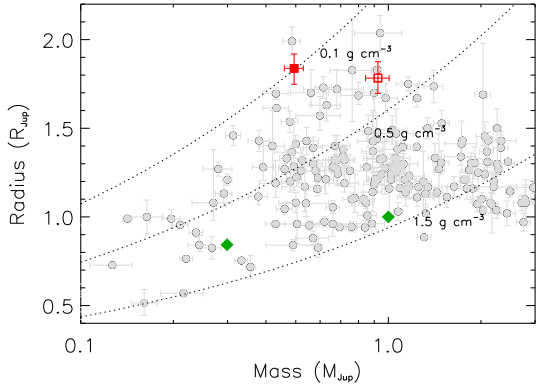}
\caption{Mass-radius diagram of the known transiting planets with $\rp \geq 0.4~\Rjup$ and 
$\mp \geq 0.1~\Mjup$. Only systems with masses determined to better than $30\%$ precision are included. 
Green diamonds indicate the Solar System giant planets Jupiter and Saturn (from right to left). 
The three dotted lines display isodensity curves of 0.1, 0.5, and 1.5~$\rm g\,cm^{-3}$. 
The position of TrES-4b as determined in this work is shown with a filled red square, to be compared with that 
(empty red square) derived by S09. 
}
\label{fig_massradius}
\end{figure}

There are several scenarios that can be proposed to explain the observed discrepancy in the RV amplitudes. 
One possible culprit might be the faint companion at $\sim1.5^{\prime\prime}$ (almost due north of TrES-4). Cunha et al. (2013) have analyzed in detail 
the impact on precision RVs of faint stellar companions from spectra gathered with fiber-fed spectrographs. The companion of TrES-4 is of late-K or 
early-M spectral type and $\approx4.5$ mag fainter at $i$-band. From Table 8 of Cunha et al. (2013) one then infers that contamination levels 
between 1 and 10 m s$^{-1}$ could apply in case the companion were to fall within the fiber of HARPS-N. A systematic effect of similar magnitude might be induced 
on the Keck RVs if the companion had been on the HIRES slit during the period of the M07 observations. Either way, this scenario does not seem to provide a 
convincing explanation for the observed difference in the $K$-value, as the higher RV dispersion inferred does not have the required magnitude and, most importantly, such 
effect would have had to occur in such a way as to exactly double (or halve) the signal amplitude. The hypothesis of large starspots on the stellar photosphere 
(causing an apparent RV shift on a timescale of the stellar rotation period) is unlikely for a late F-star such as TrES-4 (see Knutson et al. 2009). 
Large, intrinsic stellar jitter also does not appear to be supported by the observational evidence. No emission is seen in the \ion{Ca}{ii} H\&K lines related to magnetic 
activity in the HARPS-N spectra, from which we derive $\langle\log R^\prime_{HK}\rangle = -5.15\pm0.08$, essentially indistinguishable from the value reported by 
S09. The empirical relation by Wright (2005) predicts a typical stellar jitter of $\sim4$ m s$^{-1}$ for such a star as TrES-4. 
We note however that the very low value of the chromospheric emission is in line with the correlation found by Hartman (2010) 
and could be explained as the effect of absorption in the \ion{Ca}{ii} H\&K line cores by material evaporated from the low-gravity planet (Lanza 2014; Figueira et al. 2014). 
Based on the absence of bump progression in the mean line profiles of HARPS-N and archival Keck data (determined with the Donati et al. 1997 technique), we also ruled out 
the possibility that the amplitude of the RV curve of TrES-4 would be modified by non-radial stellar pulsations typical of $\gamma$ Dor variables (Kaye et al. 1999), 
which have been detected in a few cases in stars with similar stellar parameters to those of TrES-4 (Uytterhoeven et al. 2014). 
A homogeneous, comprehensive re-analysis of all available Keck data on the system might help to resolve the conundrum, particularly to understand if 
unrecognized systematics in the few Keck RVs in the discovery paper could be to blame.

\begin{figure}[t]
\vspace{-0.3cm}
\centering
\includegraphics[width=8cm]{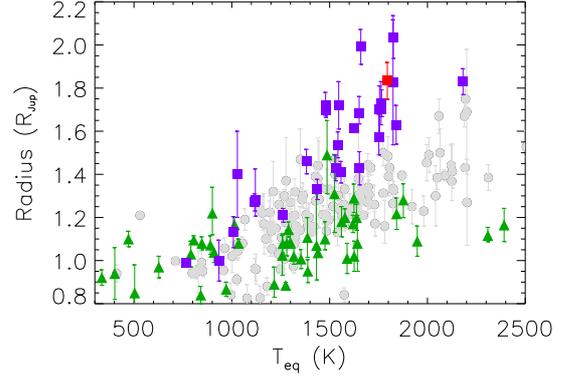}
\caption{The dependence on equilibrium temperature of the observed radii of giant planets with $\mp\geq 0.1~\Mjup$ and $\rp \geq 0.8~\Rjup$ (data from http://exoplanet.eu/). 
Purple squares, grey circles, and green triangles indicate objects with $\rho_{\rm p}\leq0.25$ g cm$^{-3}$, $0.25<\rho_{\rm p}<1.5$ g cm$^{-3}$, 
and $\rho_{\rm p}\geq1.5$ g cm$^{-3}$, respectively. The location of TrES-4b is shown with a red square.
}
\label{fig_Teqradius}
\end{figure}

The much smaller mass of TrES-4b as determined by HARPS-N implies a significantly lower density than previously thought for this hot Jupiter. 
The new location of TrES-4b in the mass-radius diagram of known transiting giant planets (Figure~\ref{fig_massradius}) makes it the second-lowest 
density object, WASP-17b (Southworth et al. 2012; Bento et al. 2013) being the record-holder at present. With a mass closer to Saturn's, TrES-4b's predicted 
radius is significantly underestimated by all empirical relations recently proposed in the literature (B\'eky et al. 2011; Enoch et al. 2011, 2012; Weiss et al. 2013), 
with radius differences ranging between 0.74~\Rjup\, and 0.45~\Rjup. In a $\rp-T_\mathrm{eq}$ diagram (see Figure~\ref{fig_Teqradius} for details) TrES-4b nicely fits in the 
upper envelope of lowest-density objects, which exhibit a strong positive correlation between the two parameters (Spearman's rank correlation coefficient $r_s=0.82\pm0.03$). 
We note how the trend of increasing $\rp$ with $T_\mathrm{eq}$ becomes significantly milder if we consider the sample of the densest giants ($r_s=0.54\pm0.06$), 
and the relationship becomes completely flat if a cut-off around 1.0~\Rjup\, is adopted (rather than the one used in Figure~\ref{fig_Teqradius}). 
 
We confirm a very low eccentricity ($e<0.016$ at the 1$\sigma$ level) for TrES-4b's orbit, improving upon the recent determination 
by K14. An estimate of the typical tidal timescales based on the model by Leconte et al. (2010) adapted as to allow constant modified 
tidal quality factors for the star ($Q^{\prime}_{*} = 10^{6}$) and the planet ($Q^{\prime}_{\rm p} = 10^{5}$) gives a circularization 
timescale of 40 Myr supporting the $e\simeq0$ hypothesis. On the other hand, the timescale for the evolution of the obliquity obtained with the same tidal model 
is $\sim 24$~Gyr, while that for the orbital decay is $\sim 6$~Gyr. This suggests that the alignment of the system is primordial and 
that no remarkable tidal evolution of the orbit has occurred during the main-sequence lifetime of the system. With the presently derived upper limit for 
the eccentricity, the maximum power dissipated by equilibrium tides inside the planet is $\sim 4.5 \times 10^{18}$~W, insufficient to explain its large radius anomaly. 
Given its peculiarity, further photometric and spectroscopic monitoring of the TrES-4 planetary system is clearly encouraged. 


\begin{acknowledgements}
The GAPS project in Italy acknowledges support from
INAF through the ''Progetti Premiali'' funding scheme of the Italian Ministry
of Education, University, and Research. We thank the TNG staff for help with the observations.
This research has made use of the results produced by the PI2S2 Project 
managed by the Consorzio COMETA, a co-funded project by the Italian Ministero dell'Istruzione, 
Universit\`a e Ricerca (MIUR) within the Piano Operativo Nazionale Ricerca Scientifica, Sviluppo Tecnologico, Alta Formazione (PON 2000-2006). 
NCS acknowledges the support from the ERC/EC under the FP7 through Starting Grant agreement n.~239953, 
from Funda\c{c}\~ao para a Ci\^encia e a Tecnologia (FCT, Portugal), and POPH/FSE (EC) through FEDER funds in
program COMPETE, as well as through and national funds, in the form of grants references RECI/FIS-AST/0176/2012 (FCOMP-01-0124-FEDER-027493), 
RECI/FIS-AST/0163/2012 (FCOMP-01-0124-FEDER-027492), and IF/00169/2012.
Operations at the Calar Alto telescope are jointly performed by the Max-Planck Institut f\"{u}r Astronomie (MPIA) and the Instituto de 
Astrof\'{i}sica de Andaluc\'{i}a (CSIC).
\end{acknowledgements}



\end{document}